\documentclass[mathleft]{an}
\usepackage{graphicx}
\usepackage{times}
\begin{document}

\Pagespan{000}{}
\Yearpublication{2007}%
\Yearsubmission{2007}%
\Month{}%
\Volume{8}%
\Issue{}%

\title{Pulsation and Orbit of AU Pegasi}

\author{M. Jurkovic,\inst{1}\fnmsep\thanks{
  \email{mojur@titan.physx.u-szeged.hu}\newline}
\ L. Szabados,\inst{2}
\ J. Vink\'{o},\inst{1}
\and B. Cs\'{a}k\inst{1}  
}
\titlerunning{Pulsation and Orbit of AU Pegasi}
\authorrunning{Jurkovic et al.}
\institute{
Department of Optics \& Quantum Electronics, University of Szeged, 
Hungary
\and 
Konkoly Observatory of the Hungarian Academy of Sciences, Budapest, 
Hungary}

\received{15 Mar 2007}
\accepted{15 May 2007}
\publonline{later}


\abstract{%
AU Pegasi is a pulsating star in a spectroscopic binary system with an 
orbital period of 53.26 days. Between 1960 and 1990 an extremely rapid period increase was observed in the value of the pulsation period, but in the last 15 years the observation show that the 
period set in 2.411 days. Fourier analysis of photometric data obtained during the ASAS project and those taken at the Piszk\'estet\H{o} 
Mountain Station of the Konkoly Observatory during 1994--2005 indicate 
that AU Pegasi is pulsating in two modes simultaneously, and the ratio of the frequencies of the two modes is 0.706, a value common for double-mode classical Cepheids. A careful analysis of other photometric observations obtained during the era of the strong period increase also revealed existence of a second mode. This may suggest that this star is not a Type II Cepheid, despite its galactic position. }

\maketitle

\section{Introduction}

AU Pegasi has been classified as a Type~II Cepheid. It is a remarkable
object among Cepheids because of its highly unstable pulsation period
(Szabados, 1977; Harris et~al., 1979) and membership in a spectroscopic
binary system with the shortest known orbital period among binaries 
involving a Cepheid component (53.3 days, Harris et~al., 1984).

The temporal behaviour of the pulsation period of \linebreak AU~Pegasi was thoroughly discussed 
by Vink\'o et~al. (1993), who followed the period changes for the interval \linebreak J.D. 2,433,100--2,448,600. 
While the pulsation period was practically constant before J.D.\,2,438,000, it was subjected to a 
strong and almost linear increase between 1964 \linebreak ($P_{\rm puls} = 2.391$ days) and 1986 
($P_{\rm puls} = 2.412$ days), which corresponds to a yearly increase of about 0.1 per cent.

The analysis of the more recent photometric data, however, indicated 
that the pulsation period became stable at the value of about 2.411 
days at the 
beginning of the 1990s. The light variations folded on this period are, 
however, 
not repetitive, which called for a deeper study of AU~Peg.

\section{Observational data and their analysis}
Photometric and radial velocity data of AU~Pegasi have been analysed. Fourier 
analysis of photometric data from All Sky Automated Survey (2003--2006) and 
from observations\linebreak made at Piszk\'estet\H{o} Mountain Station of the 
Konkoly Observatory (1994--2005) was carried out with Period04 (Lenz \& Breger, 2005). 
Radial velocity data obtained by the\linebreak Moscow CORAVEL group were taken from the papers 
by Gorynya et al. (1995, 1998). In the calculation of the orbital elements a circular orbit was 
assumed at first. The orbital parameters were obtained by iteration, in order to separate the orbital 
and pulsational velocities. 

\section{Results}
\subsection{Photometry}
\begin{figure}
\includegraphics[width=83mm,height=62mm]{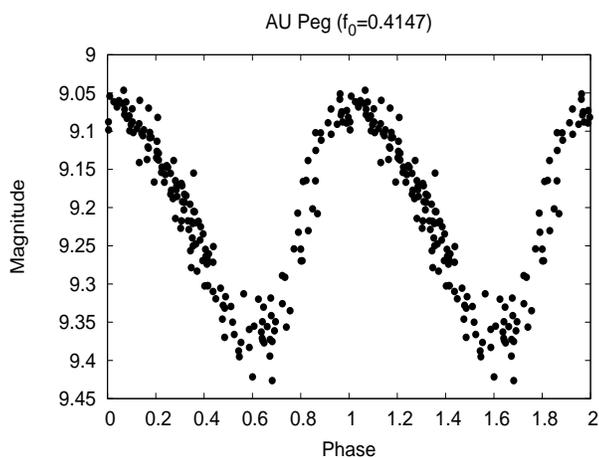}
\caption{Folded light curve of AU Peg from ASAS data}
\label{jurkovic_fig1}
\end{figure}

\begin{figure}
\includegraphics[width=83mm,height=62mm]{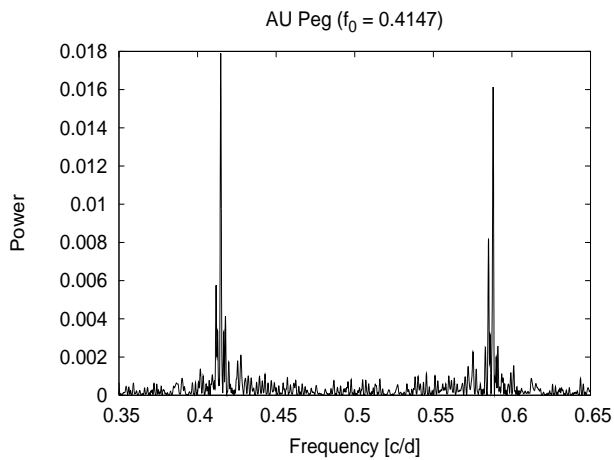}
\caption{A part of the Fourier spectrum of AU Peg obtained from the 
ASAS light curve}
\label{jurkovic_fig2}
\end{figure}

\begin{figure}
\includegraphics[width=83mm,height=62mm]{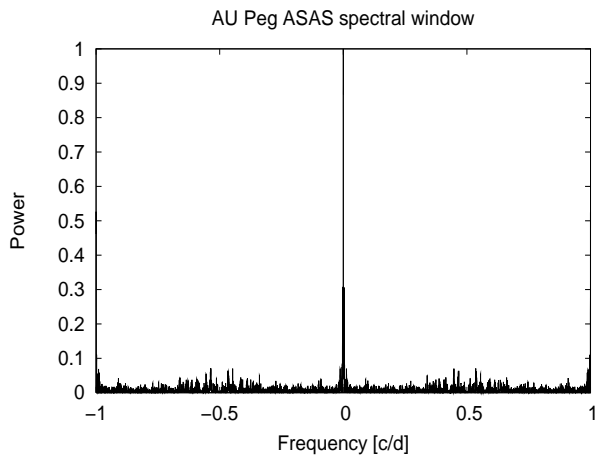}
\caption{Spectral window of ASAS data}
\label{jurkovic_fig3}
\end{figure}

In Fig. \ref{jurkovic_fig1}, we have plotted the photometric phase curve of AU Peg based 
on the data obtained during the ASAS project (Pojmanski, 2002). Fig. \ref{jurkovic_fig2} 
shows a part of its Fourier power spectrum where the spectral peak at 
$f_0=0.4147$~c/d belongs to the main pulsation frequency and the second 
highest peak is the alias of $1-f_0$. Fig. \ref{jurkovic_fig3} is the spectral window of the AU Peg ASAS data.
  
\begin{figure}
\includegraphics[width=83mm,height=62mm]{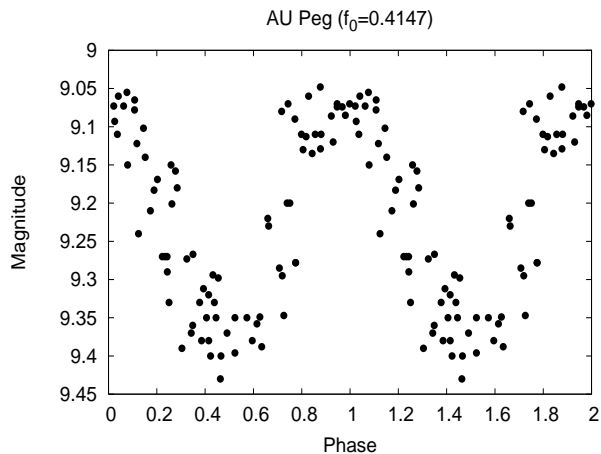}
\caption{Phased $V$ light curve of AU Peg from the Piszk\'estet\H{o} 
data}
\label{jurkovic_fig4}
\end{figure}

\begin{figure}
\includegraphics[width=83mm,height=62mm]{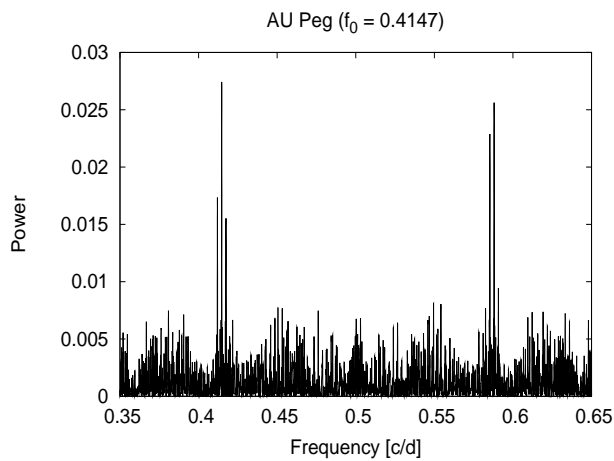}
\caption{A part of the Fourier spectrum of the Piszk\'estet\H{o} $V$ 
light curve}
\label{jurkovic_fig5}
\end{figure}

\begin{figure}
\includegraphics[width=83mm,height=62mm]{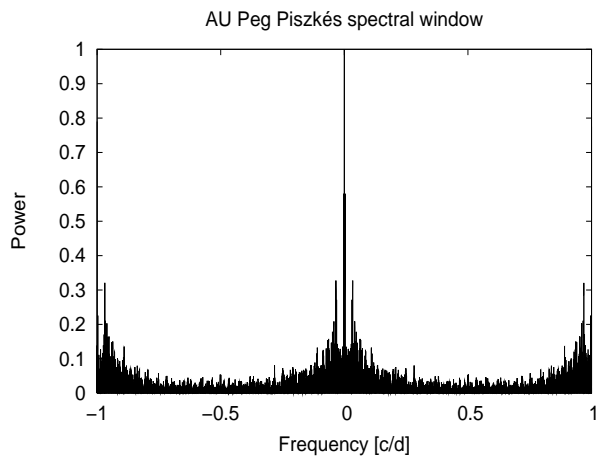}
\caption{Spectral window of Piszk\'estet\H{o} data}
\label{jurkovic_fig6}
\end{figure}

In Fig. \ref{jurkovic_fig4} and Fig. \ref{jurkovic_fig5} the folded light curve of AU Peg constructed from 
the data taken at Piszk\'estet\H{o} Mountain Station and the relevant 
part of the Fourier power spectrum of this light curve are shown. Fig. \ref{jurkovic_fig6} shows the 
spectral window. Again 
the main pulsation frequency appears at $f_0=0.4147$ accompanied by its 
alias, $1-f_0$. This pulsation frequency seems to be stable throughout 
the observation interval and corresponds to the period of 2.41138 days.

\begin{figure}
\includegraphics[width=83mm,height=62mm]{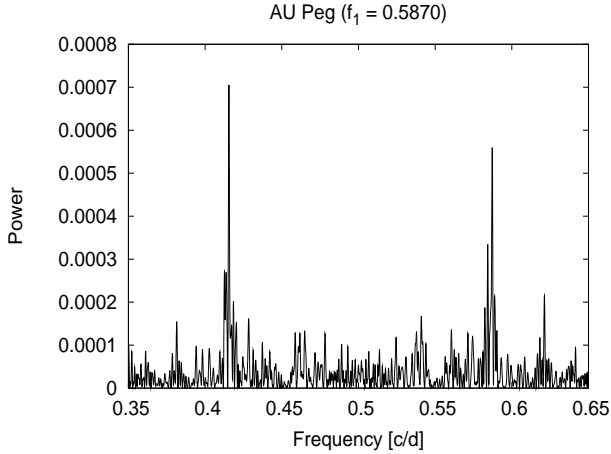}
\caption{A part of the residual Fourier power spectrum of ASAS light 
curve after prewhitening with $f_0$ and its harmonics}
\label{jurkovic_fig7}
\end{figure}

Because the scatter in the phase curve exceeds the value expected for a 
Type~II Cepheid, both data sets were analysed further. Secondary 
periodicity was detected on the residual power spectra with the frequency 
of $f_1=0.5870$ for the ASAS light curve (Fig. \ref{jurkovic_fig7}) and $f_1=0.5873$ for 
the Piszk\'estet\H{o} 
Mountain Station data (from both $V$ and $B$ photometric bands), respectively. Note that
the peak appearing at $f=0.41566$ c/d is a 1-day alias of the combination frequency $2f_0+f_1$.

The ratio of the frequencies is $f_0/f_1=0.706$ which is typical of 
double mode classical Cepheids. 
Moreover, the frequencies of the linear combinations, $f_0+f_1$, 
$f_1-f_0$, and $2f_0+f_1$ also appear in the respective Fourier spectra. From 
this we conclude that AU Peg is very likely to be a classical Cepheid 
rather than a Type II Cepheid. 

The error of the frequencies was estimated from the half width of the spectral peaks. In the case of $f_0$ and $f_1$ the error of the ASAS data is $\sigma=0.0005$, while for the Piszk\'estet\H{o} data the error is  $\sigma=0.0001$.

Other photometric data sets were also analysed in order to reveal a second pulsation mode.
Fourier analysis of photometric data from the Hipparcos 
measurements\linebreak (J.D.~2,447,889--J.D.~2,448,972, ESA 1997) showed the\linebreak 
main frequency at $f_0=0.4147$, but the data are very noisy so the 
secondary periodicity could 
not be detected. The double mode nature of AU Peg can be verified by analysing previous data published by Harris (1980), Harris et~al. (1979) and from a subsample of the observational data given by\linebreak Szabados (1977). These data
were obtained when AU Pegasi showed a significant monotonous period
increase. The individual data sets are, however, short enough to use an
appropriate `instantaneous' pulsation period when searching for 
secondary periodicity. The Fourier analysis indicates presence of double mode
pulsation in the case of all three photometric series. From the data 
obtained by Harris et~al. (1979) frequencies of $f_0=0.4162$ and $f_1=0.5911$
c/d can be deduced, Harris' (1980) data set can be well described 
by $f_0=0.4160$ and $f_1=0.5891$ c/d, while Szabados' (1977) data indicate double-mode pulsation with the frequencies $f_0=0.41652$ and $f_1=0.5898$ c/d (and their harmonics and linear combinations). From these frequencies it follows that the period of the first overtone varied simultaneously with the period of the fundamental mode, and the $f_0/f_1$ frequency ratio does not differ significantly
from the present value.

\subsection{Spectroscopy}

\begin{figure}
\includegraphics[width=83mm,height=62mm]{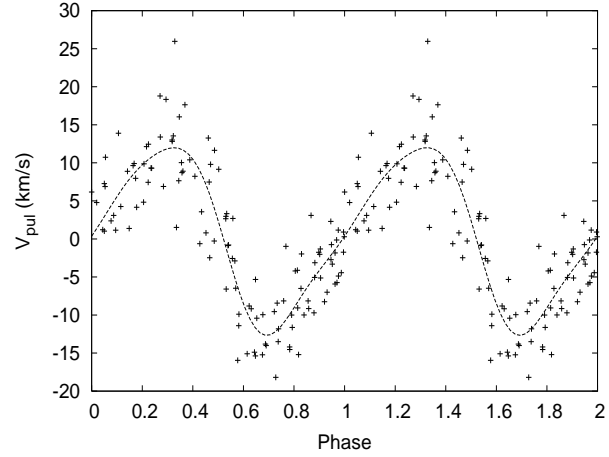}
\caption{Pulsational radial velocity curve of AU Pegasi from the Moscow 
CORAVEL data}
\label{jurkovic_fig8}
\end{figure}

\begin{figure}
\includegraphics[width=83mm,height=62mm]{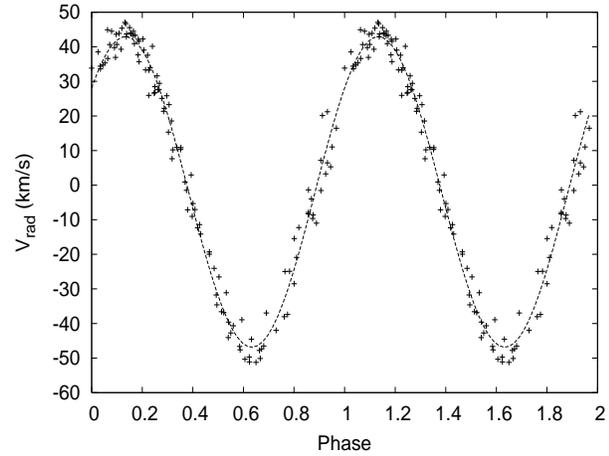}
\caption{Orbital radial velocity curve of AU Pegasi assuming a circular 
orbit}
\label{jurkovic_fig9}
\end{figure}

Spectroscopic data used in the analysis were taken from Gorynya et al. 
(1995, 1998) who published the Moscow CORAVEL radial velocities. Their radial velocity data were decomposed into the pulsational radial 
velocity curve (Fig. \ref{jurkovic_fig8}) and the orbital radial velocity curve (Fig. \ref{jurkovic_fig9}) 
-- proving the presence of the 
companion, which was earlier detected by Harris et al. (1979, 1984) and 
Vink\'{o} et al. (1993).

Assuming a circular orbit we computed the following orbital elements of the binary system:\\
$P_{\mathrm{orb}}=53.26\pm0.3 ~{\mathrm{days}}$\\
$T_0=2447739.496$\\
$\phi=-41.68\pm0.01$\\
$v_0=-1.96\pm0.42 ~{\mathrm{km\,s^{-1}}}$\\
$K=-44.86\pm0.57 ~{\mathrm{km\,s^{-1}}}$\\
$a_1\sin i=0.2196\pm0.003 ~{\mathrm{AU}}$\\
$f(m_2)=0.49\pm0.02 ~{\mathrm{M_\odot}}$.\\

We also tried to fit an elliptical orbit to the radial velocities plotted in Fig. \ref{jurkovic_fig9}. The best solution was found at $e=0.02\pm0.01$ excentricity. Therefore, the circular orbit computed above describes this 
binary system in a satisfactory manner. 

\begin{figure}
\includegraphics[width=83mm,height=62mm]{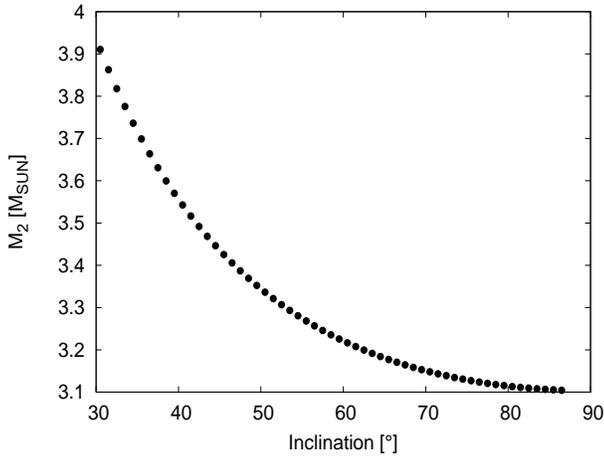}
\caption{Prediction for the mass of the companion of AU Pegasi 
according to the mass 
function $f(m_2)=0.49 ~{\mathrm{M_\odot}}$}
\label{jurkovic_fig10}
\end{figure}

The period -- radius relation for Cepheids given by\linebreak Gieren et 
al. (1998)
$$\log R=0.751(\pm0.026)\log P+1.070(\pm0.008)$$
can be used for deriving the radius of AU~Pegasi. Substituting the 
period of the fundamental mode (in days) into the above formula, we get 
that the radius of AU~Pegasi is $R=22.75\pm0.67~R_\odot$. Using the equation from 
Bono et al. (2001)  
$$\log (M_{\mathrm p}/M_\odot)=-0.09(\pm0.03)+0.48(\pm0.03)\log 
(R/R_\odot)$$
we derived that the pulsational mass for AU Pegasi is\linebreak $M_{\mathrm p}=3.64\pm0.43~M_\odot$.
Combining this result with the mass function we can calculate the mass 
of the companion as a function 
of inclination (see Fig. \ref{jurkovic_fig10}). The inclination angle must be smaller than 
$87.5^\circ$, because there is no evidence of
eclipses in the photometric data, and should not be smaller than 
$30^\circ$ because smaller values of inclination would result in physically 
unrealistic orbital radial velocity amplitude for this binary system.

\section{Summary}
\begin{itemize}
\item
Following an extended interval characterised by continuously increasing 
pulsation period, oscillations of AU Pegasi have settled at 2.411 day 
periodicity;
\item
Fourier analysis of photometric data showed that\linebreak AU~Peg is, 
in fact, a double-mode Cepheid. The frequency ratio of the two excited 
modes is $f_0/f_1=0.706$, a value typical of double-mode pulsators among 
Galactic classical Cepheids;
\item
Spectroscopic data were used to calculate the orbital elements of the 
binary system.
\end{itemize}

A more detailed discussion on the behaviour of AU~Pegasi, including the details of the period analysis of the individual data sets, the O$-$C diagram, and the results obtained from the analysis of spectroscopic data will be the topic of a forthcoming paper. 
In view of the unique properties of this Cepheid (shortest known orbital period for a Cepheid, low amplitude beat phenomenon, ambiguity in classification), AU~Pegasi deserves a closer attention from observers.

\acknowledgements
We wish to express our thanks to the organizers of the British--Hungarian--French 
N+N+N Workshop for \linebreak Young Researchers for the support given to presenting our work. This work 
has been supported by the Hungarian OTKA Grants \linebreak T~042509, TS~049872, and T~046207.


\end{document}